\documentclass[ALICE,manyauthors]{cernphprep}
\usepackage[english]{babel}
\usepackage{graphicx}
\usepackage{verbatim}
\usepackage{amsfonts}
\usepackage{makeidx}
\usepackage{color}
\usepackage{amsmath}
\usepackage{lineno}
\usepackage{epstopdf}
\usepackage{hyperref}
\newcommand{\pt}{p_\mathrm{t}} 

\begin{document}
\begin{titlepage}

\PHnumber{2011-196}                 
\PHdate{27 November 2011}                 

\title{Light vector meson production in pp collisions at $\sqrt{s}$ = 7 TeV}

\Collaboration{ALICE Collaboration%
         \thanks{See Appendix~\ref{app:collab} for the list of collaboration 
                      members}}

\ShortAuthor{ALICE Collaboration}
\ShortTitle{Light vector meson production in pp collisions at $\sqrt{s}$ = 7 TeV}

\begin{abstract}
\label{abstract}
The ALICE experiment has measured low-mass dimuon production 
in pp collisions at $\sqrt{s}$ = 7 TeV in the dimuon rapidity 
region $2.5<y<4$. The observed dimuon mass spectrum is described 
as a superposition of resonance decays ($\eta, \rho, \omega,
\eta^{'}, \phi$) into muons and  
semi-leptonic decays of charmed mesons. 
The measured production cross sections for $\omega$ and $\phi$ 
are $\sigma_\omega (1<\pt<5~{\rm GeV}/c,2.5<y<4) = 5.28 \pm 0.54 \mathrm{(stat)} \pm 0.50 \mathrm{(syst)~mb}$
and $\sigma_\phi(1<\pt<5~{\rm GeV}/c,2.5<y<4)=0.940 \pm  0.084 (\mathrm{stat}) \pm 0.078 \mathrm{(syst)~mb}$.
The differential 
cross sections ${\rm d}^2\sigma/{\rm d}y {\rm d}\pt$
are extracted as a function of $\pt$ for $\omega$ and $\phi$. The 
ratio between the $\rho$ and $\omega$ cross section is obtained. 
Results for the $\phi$ are compared with other measurements at the 
same energy and with predictions by models. 
\end{abstract}

\end{titlepage}
\setcounter{page}{2}
%
\section{Introduction}
\label{Introduction}
The measurement of light vector meson production ($\rho, \omega, \phi$) 
in pp collisions provides insight into soft Quantum Chromodynamics (QCD) processes 
in the LHC energy range. 
Calculations in this regime are based on QCD inspired 
phenomenological models~\cite{SjostrandSkands} that must be tuned to the data, in
particular for hadrons that contain the $u$, $d$, $s$ quarks. 
The evolution of particle production as a function of $\sqrt{s}$ is difficult 
to estabilish. 
Measurements at mid-rapidity in pp collisions at the beam injection energy of the LHC 
($\sqrt{s}=0.9$~TeV) were performed by the ALICE experiment~\cite{ALICEStrangeness}, 
and compared with several PYTHIA~\cite{PYTHIA} tunes and PHOJET~\cite{PHOJET}. The 
comparison showed that, for transverse momenta larger than $\sim 1$~GeV/$c$,
the strange particle spectra are strongly understimated by the models, 
by a factor of 2 for $\textrm{K}^0_S$ and 3 for hyperons, with a 
smaller discrepancy for the $\phi$. Extending the measurements to 
larger energies and complementary rapidity domains is needed in order 
to further constrain the models.

Moreover, light vector meson production provides a reference 
for high-energy heavy-ion collisions. 
In fact, key information on the hot and dense state of strongly 
interacting matter produced in these collisions can be extracted
measuring light meson production. 

The ALICE experiment at the LHC can access vector mesons produced
in the rapidity range $2.5<y<4$ through their decays 
into muon pairs~\footnote{In the ALICE coordinates,
the muon spectrometer covers the pseudo-rapidity range
$-4<\eta<-2.5$, where the $z$ axis is oriented along 
the beam direction. However, since in pp collisions 
results are symmetric with respect to $y=0$, we prefer to drop the 
negative sign when quoting the rapidity values.}. 
In this Letter we report results obtained in pp collisions 
at $\sqrt{s}=7$~TeV in the dimuon transverse momentum range $1 < \pt < 5$~GeV/$c$ 
based on the full data sample collected in 2010 with a minimum bias muon trigger. 
The measurement is done via a combined fit of the dimuon invariant 
mass spectrum after combinatorial background subtraction. 
%
%
\section{Experimental setup} 
\label{Experimental setup} 
The ALICE detector is fully described elsewhere~\cite{ALICE}. 
The main detectors relevant for this analysis are the forward muon spectrometer,
which covers the pseudo-rapidity region $-4 < \eta < -2.5$, the VZERO detector and 
the Silicon Pixel Detector (SPD) of the Inner Tracking System.
  
The elements of the muon spectrometer are a front hadron absorber, 
followed by a set of tracking stations, a dipole magnet, an iron wall acting
as muon filter and a trigger system. 
 
The front hadron absorber  is made of carbon, concrete and steel and is placed 
at a distance of 0.9~m from the nominal interaction point (IP). 
Its total length of material corresponds to ten hadronic interaction lengths. 
The dipole magnet is 5~m long and 
provides  a magnetic field of up to $0.7$~T in the vertical direction 
which gives a field integral of $3~\mathrm{Tm}$.
 
The muon tracking is provided by a set of five tracking stations, each one composed of 
two cathode pad chambers. The stations are located between 5.2 and 14.4~m from 
the IP, the first two upstream of the dipole magnet, the third in the middle 
of the dipole magnet gap and the 
last two downstream.  The intrinsic spatial resolution of the tracking chambers is 
$\sim 100~\mu\mathrm{m}$ in the bending direction. 
 
A 1.2~m thick iron wall, corresponding to 7.2 hadronic interaction lengths, 
is placed between the tracking and trigger systems and  
absorbs the residual secondary hadrons emerging from the front absorber. 
The front absorber together with the muon filter stops muons with momentum lower than 4~GeV/$c$.
The muon trigger system consists of two detector stations, placed at 16.1 and 
17.1~m from the IP. Each one is composed of two planes 
of resistive plate chambers (RPC), with a time resolution of about 2~ns. 
 
 
The SPD consists of two cylindrical layers of silicon pixel detectors, 
positioned at a radius of 3.9 and 7.6~cm from the beam. The pseudo-rapidity range covered by the
inner and the outer layer is $|{\eta}|<2.0$ and $|{\eta}|<1.6$, respectively. 
Besides contributing to the 
primary vertex determination, it is used for the input of the 
level-0 trigger (L0). 
 
The VZERO detector consists of two arrays of plastic scintillators
placed at 3.4~m and -0.9~m from the IP and covering the pseudo-rapidity regions 
$2.8<\eta<5.1$ and $-3.7<\eta<-1.7$, respectively.
This detector provides timing information for the L0 trigger
and has a time resolution better than 1~ns, thus giving the possibility to
reject beam-halo and beam-gas interactions in the off-line analysis.
 
%
%
\section{Data selection and analysis}
\label{Data selection and analysis}
During the pp run in 2010, the instantaneous luminosity delivered by the LHC to ALICE 
ranged from $0.6 \times 10^{29}$ to 
$1.2 \times 10^{30} \mathrm{cm}^{-2} \mathrm {s}^{-1}$.
The fraction of events with multiple interactions in a single
bunch crossing was less than $5\%$.
The data sample used in this analysis was collected using the 
muon trigger, which is activated when at least three 
of the four RPC planes in the two muon trigger stations 
give a signal compatible with 
a track in the muon trigger system. 
To evaluate the integrated luminosity ($\mathcal{L}_{\rm int}$), 
a minimum bias (MB) trigger, independent of the muon trigger, 
was collected in parallel. It is activated when at least 
one out of the 1200 SPD readout chips detects a hit or when at least 
one of the two VZERO scintillator arrays has fired, in coincidence 
with the arrival of bunches from both sides.

The integrated luminosity was determined by measuring 
the MB cross section $\sigma_{\mathrm{MB}}$
and counting the number of MB events. 
The $\sigma_{\mathrm{MB}}$ value is 62.3~mb, and is affected by a 
$4\%$ systematic uncertainty. It was obtained measuring the  
cross section $\sigma_{\mathrm{V0AND}}$~\cite{Gagliardi}, for the 
occurrence of coincident signals in the two VZERO detectors (V0AND) 
in a van der Meer scan~\cite{vdm}. 
The factor $\sigma_{\mathrm{V0AND}}/\sigma_{\mathrm{MB}}$ was obtained 
as the fraction of MB events where the L0 trigger input corresponding 
to the V0AND condition has fired. Its value is 0.87 and is stable 
within $0.5\%$ over the analyzed data.
The full data sample used for this analysis, amounting to 
an integrated luminosity of approximately 85~nb$^{-1}$, 
was used to extract the $\pt$ distributions. 
Part of the data was not collected with the MB trigger in parallel 
with the muon trigger. For this fraction, the integrated 
luminosity could not be measured and the $\omega$ and $\phi$ cross sections were determined 
with the remaining subsample corresponding to $\mathcal{L}_{\rm int}=55.7~\mathrm{nb}^{-1}$.
 
Track reconstruction in the muon spectrometer is based on a Kalman filter 
algorithm~\cite{OfflineNote,kalmanC}. 
Straight line segments are formed from the clusters on the two planes of each
of the most downstream tracking stations (4 and 5), since these are less affected by 
the background coming from soft particles that emerge from the front absorber. 
Track properties are first estimated assuming that tracks originate from 
the IP and are bent in a uniform magnetic field in the dipole. Afterwards, 
track candidates starting in station 4 are extrapolated to station 5, or vice versa, 
and paired with at least one cluster on the basis of a $\chi^2$ cut. Parameters 
are then recalculated using the Kalman filter. The same procedure is applied 
to the upstream stations, rejecting track candidates that cannot be matched 
to a cluster in the acceptance of the spectrometer.
Finally, fake tracks that share the same cluster with other tracks are removed and 
a correction for energy loss and multiple Coulomb scattering in the absorber is 
applied by using the Branson correction~\cite{OfflineNote}. 
The relative momentum resolution of the reconstructed tracks ranges 
from 1$\%$ at 20~GeV/$c$ to 4$\%$ at 100~GeV/$c$. 

Muons were selected requiring that the direction and position of
each muon track reconstructed in the tracking chambers match the ones of the 
corresponding track in the trigger stations. A cut on the muon 
rapidity $2.5<y_{\mu} < 4$ was applied in order to remove the tracks 
close to the acceptance borders. Muon pairs were selected requiring that 
both muons satisfy these cuts. 
Approximately 291,000 opposite-sign $(N_{+-})$ and 
197,000 like-sign $(N_{++},~N_{--})$ muon pairs passed these selections. 
 
\begin{figure}[tb]
	\centering
	\includegraphics[width=0.99 \columnwidth]{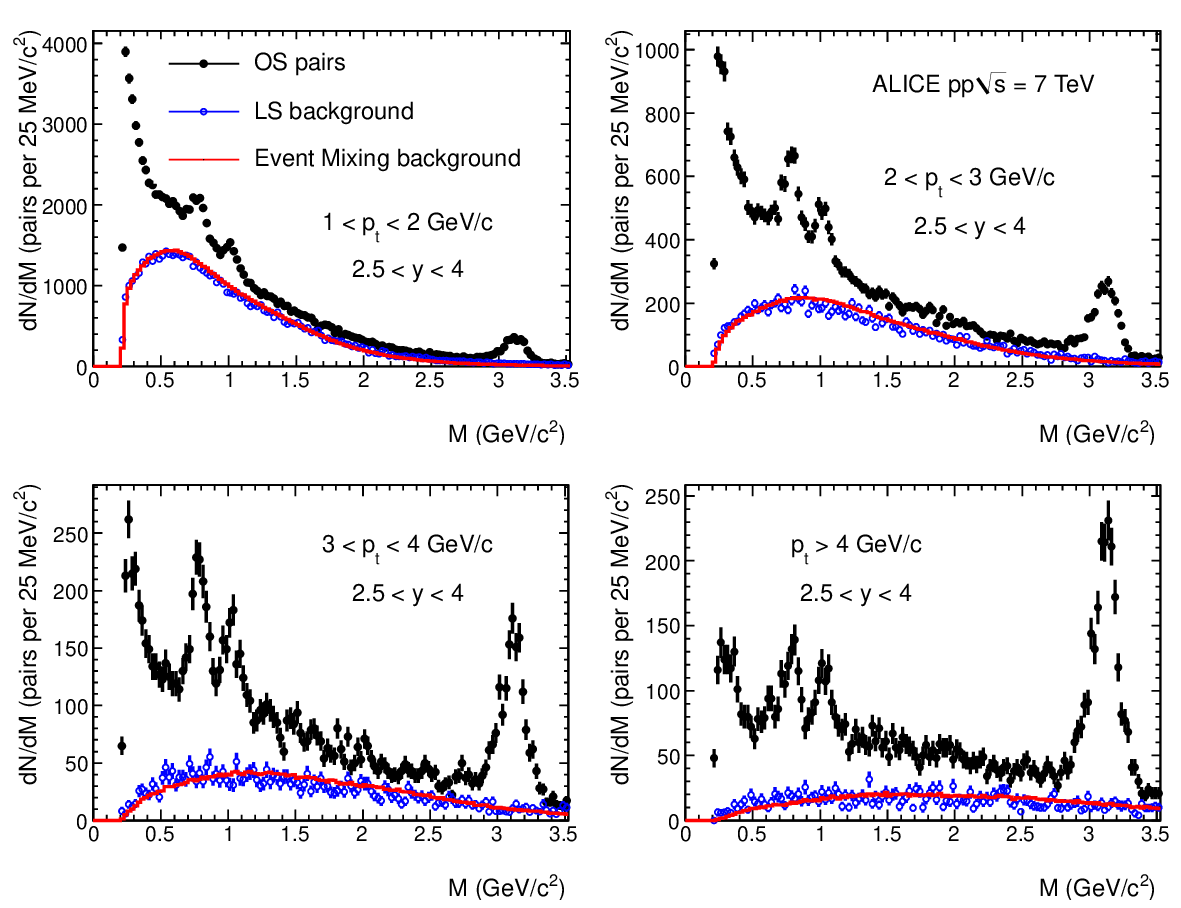}
	\caption{(Color online) Invariant mass spectra for opposite-sign muon pairs in pp at 
			 $\sqrt{s}=7$~TeV in different $\pt$ ranges. 
			 The combinatorial background, evaluated from opposite-sign 
			 pairs in mixed events (red line) or like-sign pairs in non-mixed events 
			 (blue points), is also shown.}
\label{fig:m_os_vs_pt}
\end{figure}

The opposite-sign pairs are composed of correlated and uncorrelated 
pairs. The former constitute the signal, while the latter, coming 
mainly from decays of pions and kaons into muons, 
form the combinatorial background, which was evaluated using an 
event mixing technique.
The distribution obtained was normalized to $2 R \sqrt{N_{++}N_{--}}$, 
where $N_{++}~(N_{--})$ is the number of like-sign positive (negative) 
pairs integrated in the full mass range.
It is assumed that the like-sign pairs 
are uncorrelated. The fraction of correlated like-sign pairs, coming from 
the decay chain of beauty mesons and $B-\bar B$ oscillations~\cite{crochetPBM} 
was determined from the measured open charm content and the ratio between open beauty 
and charm (see below). It amounts to $\approx 0.5 \%$ for 
$\pt > 1$~GeV/$c$ and $M<1.5$~GeV/$c^2$, and was thus neglected. 
The $R$ factor is defined as $A_{+-}/\sqrt{A_{++}A_{--}}$, 
where $A_{++} (A_{--})$ is the acceptance for a $++$ ($--$) pair, and 
takes into account possible correlations introduced by the detector. 
It was evaluated using two methods. The first employs MC
simulations to determine the acceptances $A_{\pm\pm}$.
The other method uses the mixed-event pairs to estimate $R$ as 
$R=N^{\mathrm{mixed}}_{+-}/\sqrt{N^{\mathrm{mixed}}_{++} N^{\mathrm{mixed}}_{--}}$, 
where $N^{\mathrm{mixed}}_{\pm \pm}$ is the number of mixed 
pairs for a given charge combination. The two methods are in agreement 
for $\pt>1$~GeV/$c$. We obtain $R=0.95$ for $\pt>1$~GeV/$c$.  
The event mixing procedure was cross-checked by comparing the results obtained for 
like-sign mixed pairs with the non-mixed ones. The shapes are identical, 
while the number of like-sign pairs estimated with the 
event mixing is lower than the one in the data by $5\%$. We take 
this value as the systematic uncertainty on the background normalization.  
The signal-to-background ratio for $\pt>1$~GeV/$c$ is 
about 1 at the $\phi$ and $\omega$ masses.
Alternatively, the combinatorial background can be evaluated 
using only the like-sign pairs in the non-mixed data, and calculating 
for each $\Delta M$ mass bin the quantity 
$2 R(\Delta M) \sqrt{N_{++}(\Delta M) N_{--} (\Delta M)}$. 
Figure~\ref{fig:m_os_vs_pt} shows the invariant mass spectrum for 
opposite sign muon pairs in different $\pt$ ranges, together with 
the combinatorial background estimated with the event mixing technique 
or using the like-sign pairs. It is seen that the two techniques 
are in good agreement for $\pt > 1$~GeV/$c$. For lower pair transverse 
momenta both methods fail in describing the background. In this region, 
the method based on the like-sign pairs gives a backgound mass spectrum 
that overshoots the opposite-sign pair spectrum, while the event mixing 
technique does not reproduce the non-mixed like-sign pairs spectra. 
The analysis is thus limited to $\pt>1$~GeV$/c$. The event mixing 
technique is used, since it is less affected by statistical fluctuations.
%
%
%
\begin{figure}[tb]
	\centering
	\includegraphics[width=0.60 \columnwidth]{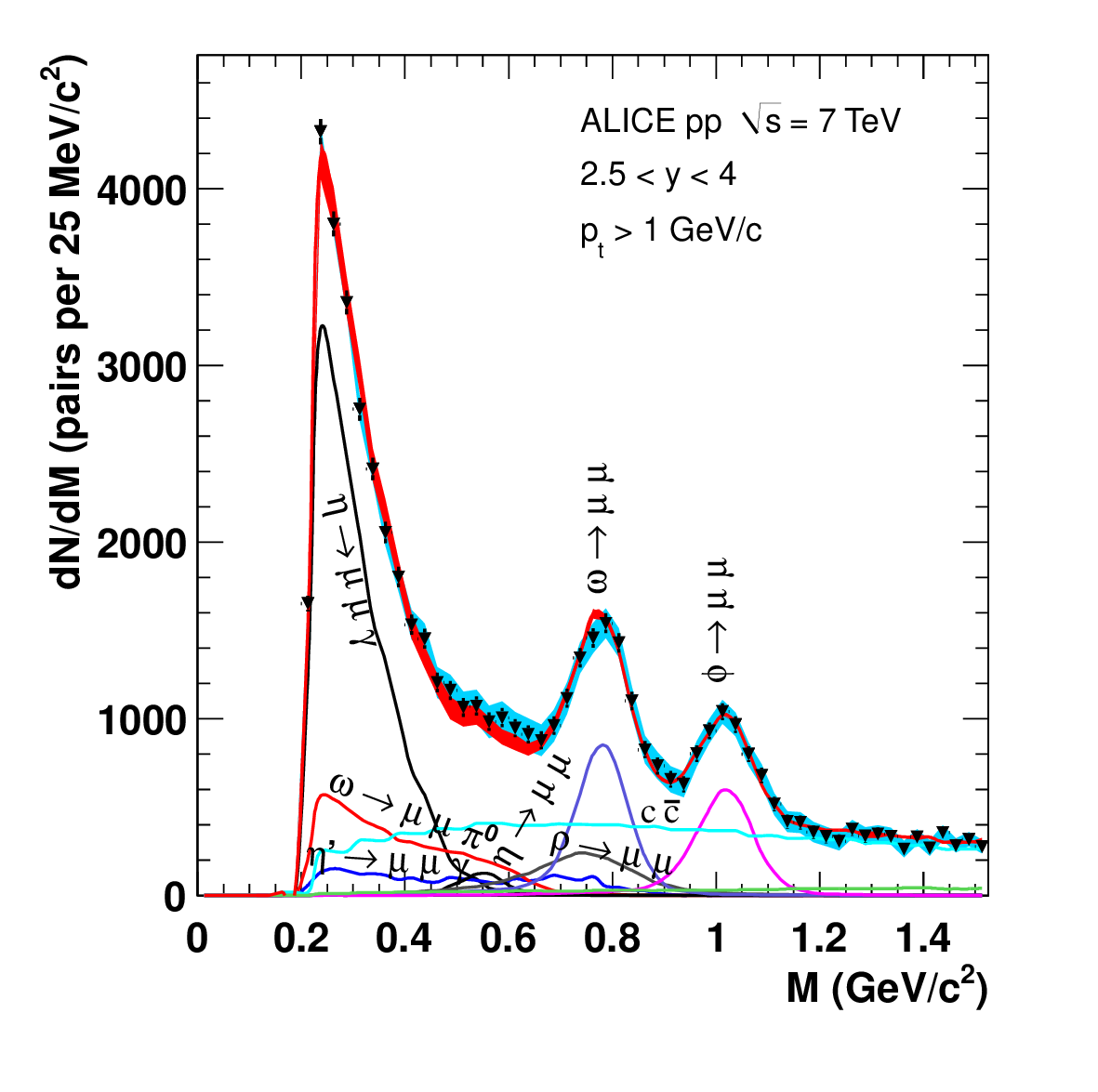}
	\caption{(Color online) Dimuon invariant mass spectrum in pp at $\sqrt{s}=7$~TeV 
			 after combinatorial background subtraction for $\pt>1$~GeV$/c$ (triangles).
   	         Light blue band: systematic uncertainty from background subtraction. 
	         Red band: sum of all simulated contributions. The width of the 
	         red band represents the uncertainty on the relative normalization of the sources. 
	}
\label{fig:mass_spectrum}
\end{figure}
%

After subtracting the combinatorial background from the opposite-sign mass spectrum, 
we obtain the raw signal mass spectrum shown in Fig.~\ref{fig:mass_spectrum}. 
The mass resolution at the $\phi$ mass is $\sigma_M \approx 60$~MeV/$c^2$, in good 
agreement with the Monte Carlo simulation. 
The processes contributing to the dimuon mass spectrum are 
the light meson ($\eta$, $\rho$, $\omega$, $\eta^{'}$, $\phi$) 
decays into muons and the correlated semi-leptonic open charm and beauty decays. 
The light meson contributions were obtained performing a simulation based on a hadronic 
cocktail generator. The input rapidity distributions for all particles are 
based on a parametrization of PYTHIA 6.4~\cite{PYTHIA} results obtained with the 
Perugia-0 tune~\cite{Perugia0}. The same procedure is followed for the $\eta'$ 
$\pt$ distribution, while for $\rho$, $\omega$ and $\phi$ the transverse momentum
is described with a power-law function, used also by the HERA-B experiment to fit 
the $\phi$ $\pt^2$ spectrum~\cite{HERAB}: 
\begin{equation}
\label{powerlaw}
{{\mathrm{d} N} \over {\mathrm{d} \pt}} = C {{\pt} \over {[1 + (\pt /p_0)^2]^n}}.
\end{equation}
The parameters $n$ and $p_0$ were tuned iteratively to the results of 
this analysis. 
The $\pt$ distribution of $\eta$ is based on preliminary results from  
$\eta$ production yields measured in the two-photon decay channel by ALICE~\cite{marin}. 
The open charm and beauty generation 
is based on a parameterization of PYTHIA~\cite{OfflineNote}. The detector response 
for all these processes is obtained with a
simulation that uses the GEANT3~\cite{geant} transport code. 
The simulation results are then subjected to the same reconstruction 
and selection chain as the real data. 
The invariant mass spectrum is fitted with a superposition of the
aforementioned contributions. 
The free parameters of the fit are the normalizations of 
the $\eta\rightarrow\mu \mu \gamma$, $\omega\rightarrow\mu \mu$, 
$\phi\rightarrow\mu \mu$ and open charm signals. The
processes $\eta \rightarrow \mu\mu$ and $\omega \rightarrow \mu\mu \pi^0$
are fixed according to the relative branching ratios. 
The contribution from $\rho \rightarrow \mu\mu$ 
was fixed by the assumption that the production cross section of 
$\rho$ and $\omega$ are equal~\cite{NA27,CERESpA,NA60pA}. The $\eta'$ contribution was set fixing the 
ratio between the $\eta'$ and $\eta$ cross sections according to 
PYTHIA. The ratio between the open beauty and open charm was 
fixed according to the results from the LHCb Collaboration~\cite{LHCbcharm,LHCbBeauty}.
The main sources of systematic uncertainty are the 
background normalization and the relative normalization of the sources, 
mainly due to the error on the branching ratios for
the $\omega$ and $\eta'$ Dalitz decays.
The raw numbers of $\phi$ and $\rho+\omega$ resonances obtained from the fit are
$N_\phi^\mathrm{raw}=(3.20 \pm 0.15) \times 10^3$ and 
$N_{\rho+\omega}^\mathrm{raw}=(6.83 \pm 0.15) \times 10^3$. 
%
%
\section{Results}
\label{Results}
The $\phi$ production cross section was evaluated in the range 
$2.5 < y < 4$, $1 < \pt < 5$~GeV/$c$ through the formula:
$$\sigma_\phi =
{N_\phi^\mathrm{raw} \over {A_\phi \varepsilon_\phi BR(\phi \rightarrow l^+ l^-)}}
{\sigma_{MB} \over N_{MB}} 
{N^{MB}_\mu \over N^{\mu-MB}_\mu},$$ 
where $N_\phi^\mathrm{raw}$ is the measured number of $\phi$ mesons, 
$A_\phi$ and $\varepsilon_\phi$  are the geometrical acceptance 
and the efficiency respectively, $N_{MB}$ is the number of minimum bias collisions,
$\sigma_{MB}$ is the ALICE minimum bias cross section in pp collisions
at $\sqrt{s}=7$~TeV, and 
${N^{MB}_\mu / N^{\mu-MB}_\mu}$ is the ratio between the number 
of single muons collected with the minimum bias trigger and 
with the muon trigger in the region $2.5 < y_\mu < 4$, 
$\pt> 1$~GeV/$c$. The number of minimum bias collisions was 
corrected, as a function of time, by the probability to have multiple interactions
in a single bunch crossing. Finally,
$BR(\phi \rightarrow l^+ l^-) = (2.95 \pm 0.03) \times 10^{-4}$
is the branching ratio into lepton pairs. Assuming lepton universality, 
this number is obtained as a weighted mean of the measured branching 
ratio in $\mu^+\mu^-$ with that into $\rm e^+e^-$, 
because the latter has a much smaller 
experimental uncertainty than the former~\cite{PDG}.  
The number of $\phi$ mesons was evaluated by performing a fit to the mass
spectrum for each $\Delta \pt=0.5$~GeV/$c$ interval in the transverse 
momentum range covered by the analysis. The acceptance-corrected results 
were then summed in order to obtain the total number of $\phi$ mesons.
In this way the dependence of the acceptance correction on the input
$\pt$ distribution used for the Monte Carlo simulation becomes insignificant. 
Alternatively, a fit was performed on the mass spectrum integrated over
 $1 < \pt < 5$~GeV/$c$ and a global correction factor was applied. 
The results of the two approaches agree within 3$\%$. The first 
approach was used for the results reported in this paper. 
The $\phi$ meson acceptance and efficiency correction in the range covered 
by this analysis was evaluated through Monte Carlo simulations and 
ranges from $10\%$ to $13\%$, depending on the data-taking period. 
The ratio ${N^{MB}_\mu / N^{\mu-MB}_\mu}$ strongly depends on the 
data taking conditions and was evaluated as a function of time. 

%
\begin{figure}[tb]
	\centering
	\includegraphics[width=0.60 \columnwidth]{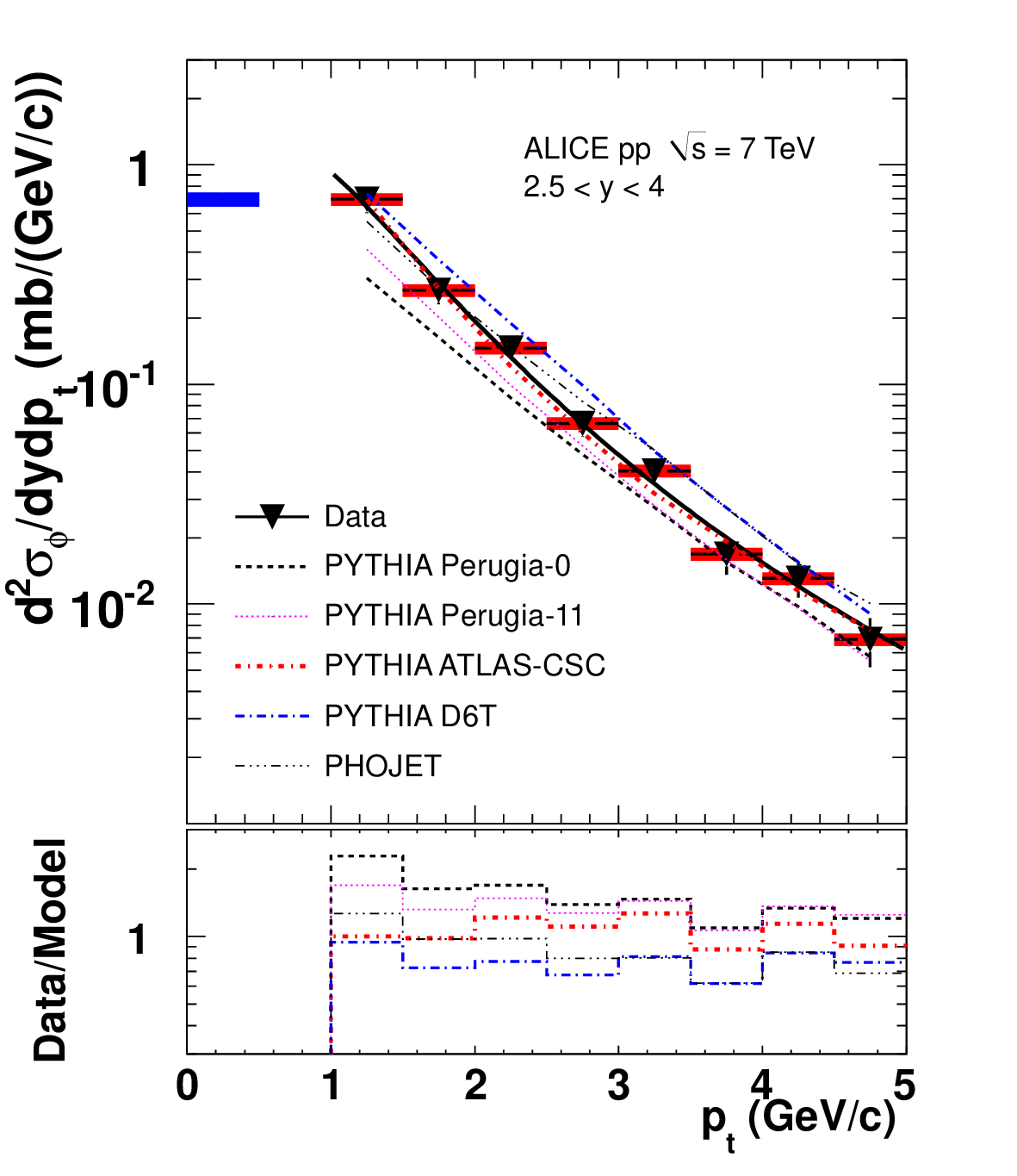}
	\caption{Top: Inclusive differential $\phi$ production cross section
	${\rm d}^2\sigma_\phi/{\rm d}y {\rm d}\pt$ for $2.5 < y < 4$. 
	The error bars represent the quadratic sum of the statistical and systematic
	uncertainties, the red boxes the 
	point-to-point uncorrelated systematic uncertainty, the blue box on 
	the left the error on normalization. Data are fitted with Eq.~(\ref{powerlaw})
	(solid line) and compared with the Perugia-0, Perugia-11, 
	ATLAS-CSC and D6T PYTHIA tunes and with PHOJET.
	Bottom: Ratio between data and models. 	
	} 
\label{fig:dsigmaphidydpt}
\end{figure}
%

We obtain 
$\sigma_\phi(1<\pt<5~{\rm GeV}/c,2.5<y<4)=0.940 \pm  0.084 (\mathrm{stat}) \pm 0.078 \mathrm{(syst)~mb}$.
The systematic uncertainty results from the uncertainty on the background subtraction
($2 \%$), the $\phi$ branching ratio into dileptons ($1\%$), 
the muon trigger and tracking efficiency 
($4\%$ and $3\%$ respectively),
the minimum bias cross section ($4\%$) 
and the ratio ${N^{MB}_\mu / N^{\mu-MB}_\mu}$ ($3\%$).
The first two contributions have been described above.
The others are common 
to all analyses in the dimuon channel, and are extensively 
discussed elsewhere~\cite{jpsipaper}. Here, only the main points are 
briefly summarized.
The muon trigger efficiency was estimated measuring the number of 
$J/\psi$ mesons decaying into muons, after efficiency and acceptance
corrections, in two ways: in the first case both muons 
were required to match the trigger, while in the second only one 
muon needed to fulfill this condition.
The tracking efficiency was evaluated starting from the determination of 
the efficiency for individual chambers, computed by taking advantage from
the redundancy of the tracking information in each station. 
The same procedure was applied to the data and to the Monte Carlo 
simulations. The differences in the results give the systematic 
uncertainty on the tracking efficiency. 
The error on the minimum bias cross section is 
mainly due to the uncertainties in the beam intensities~\cite{beamIntensity} 
and in the analysis procedure adopted for the determination of the beam luminosity 
via the van der Meer scan. 
The error on the ratio ${N^{MB}_\mu / N^{\mu-MB}_\mu}$ was evaluated 
comparing the value measured as described above with the information 
obtained from the trigger scalers, taking into account the dead time of the triggers~\cite{lumiScalers}. 

Table~\ref{tab:compModels} compares the present measurement 
with some commonly used tunes of PYTHIA~\cite{PYTHIA} 
(Perugia-0~\cite{Perugia0}, Perugia-11~\cite{Perugia11}, ATLAS-CSC~\cite{ATLAS} and D6T~\cite{D6T}) 
and PHOJET~\cite{PHOJET}. It can be seen that Perugia-0 and Perugia-11
underestimate the $\phi$ cross section (by about a factor of 2 and 
1.5, respectively), while the others agree with the measurement within its error. 

%
\begin{figure}[tb]
	\centering
	\includegraphics[width=0.60 \columnwidth]{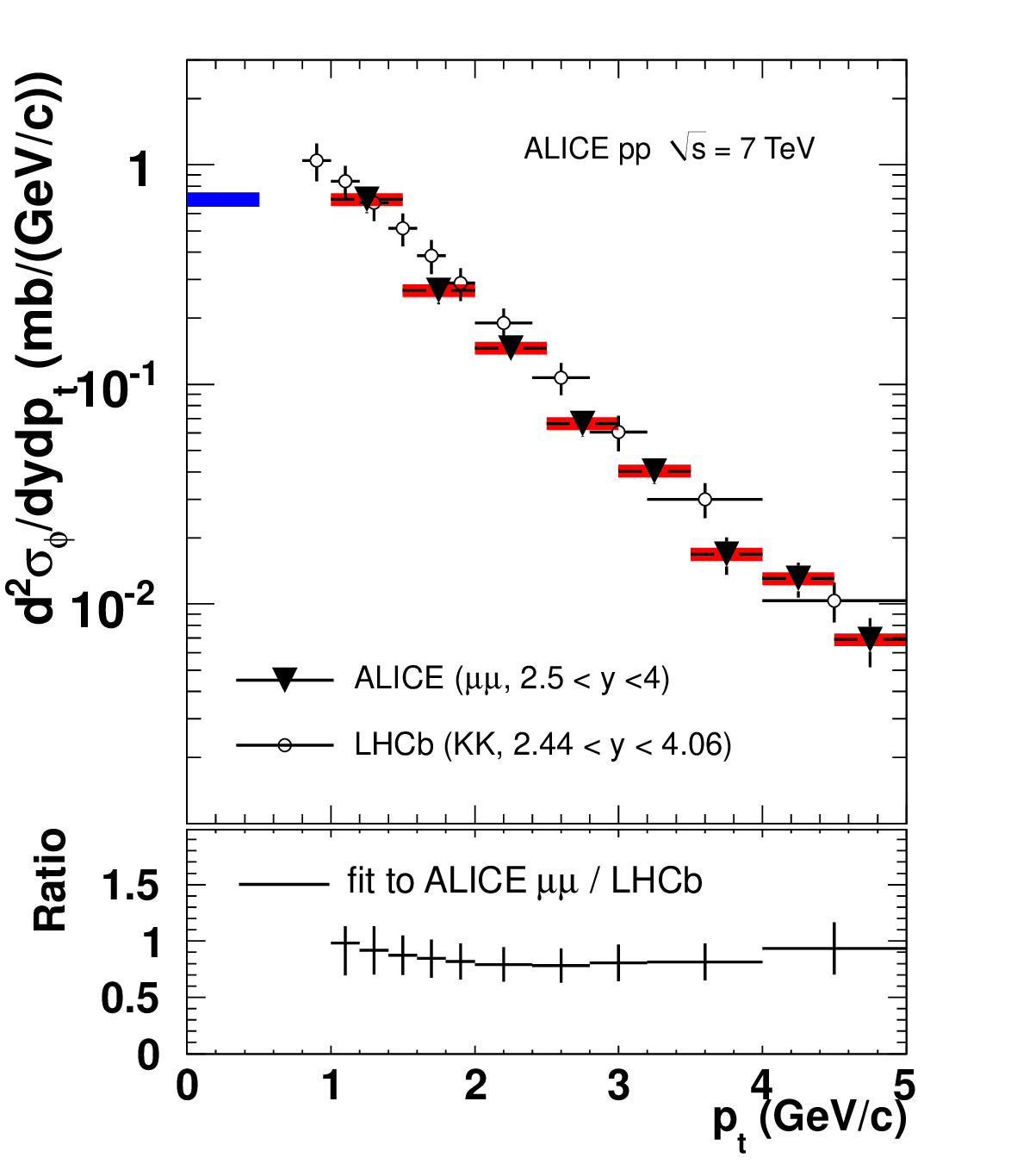}
	\caption{(Color online) Top: Inclusive differential $\phi$ production cross section
	${\rm d}^2\sigma_\phi/{\rm d}y {\rm d}\pt$, as measured via the decay 
	into dimuons (black triangles). The blue box on 
	the left represents the error on normalization.
	The data are compared to the measurements 
	in the kaon decay channel by LHCb (black open circles)~\cite{LHCb}.
	Bottom: Fit to the differential cross section measured in dimuons 
	divided by the cross section measured in the kaon channel by LHCb. 
	}
\label{fig:phi_all}
\end{figure}
%
The differential cross section ${\rm d}^2\sigma_\phi/{\rm d}y {\rm d}\pt$ 
is shown in Fig.~\ref{fig:dsigmaphidydpt} (top). 
Numerical values are reported in Table~\ref{tab:xsect}. 
$\pt$-dependent contributions to the 
systematic uncertainties, due to the uncertainty on trigger and tracking efficiency and
background subtraction, are indicated as red boxes. 
The uncertainty on the minimum bias cross section, 
branching ratio and ${N^{MB}_\mu / N^{\mu-MB}_\mu}$ ratio contribute to the 
uncertainty in the overall normalization. 
As stated above, the $\phi$ cross section is extracted from a subsample of the 
data used to determine the $\pt$ distribution, and is thus affected by a larger 
statistical uncertainty, resulting in a $5\%$ contribution to the normalization error.
Fitting the expression in Eq.~(\ref{powerlaw}) (solid line) to the differential cross section gives
$p_0=1.16 \pm 0.23$~GeV/$c$ and $n=2.7\pm 0.2$. The PYTHIA and PHOJET 
predictions are also displayed in Fig.~\ref{fig:dsigmaphidydpt}, where the bottom panel shows the 
ratio between the measurement and the model predictions. PYTHIA with the ATLAS-CSC 
and D6T tunes reproduces the measured differential cross section, while the others predict
a harder $\pt$ spectrum. 

The results are compared to measurements of $\phi \rightarrow \mathrm{K}^+\mathrm{K}^-$ for $2.44<y<4.06$ 
by the LHCb Collaboration~\cite{LHCb} in Fig.~\ref{fig:phi_all}.
The observed shapes of the $\pt$ distributions are similar.
In order to compare with our integrated cross section result, 
the differential cross section measurement by LHCb was integrated for $\pt>1$~GeV$/c$
and scaled by a small correction factor, obtained from PYTHIA (Perugia-0), to account 
for the slight difference in rapidity acceptance. The result is 
$\sigma_\phi=1.07\pm 0.15(\mathrm{stat.+syst.})$~mb.
When the statistical errors and the part of the systematic uncertainty
which is not correlated among the two experiments are properly taken into account,
the two measurements are in agreement. 
%
\begin{figure}
	\centering
	\includegraphics[width=0.60 \textwidth]{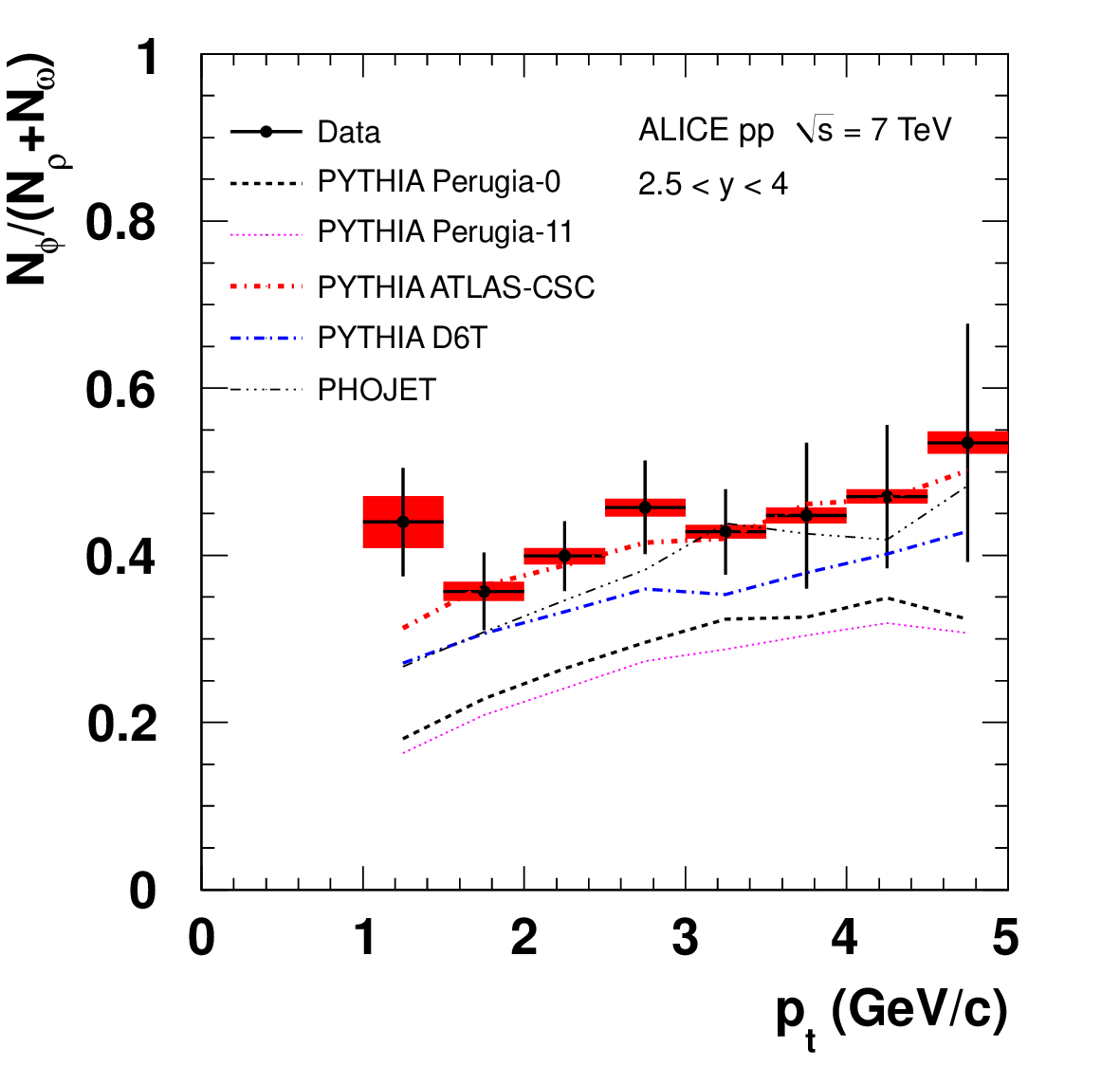}
	\caption{Ratio $N_\phi /(N_\rho + N_\omega)$ as a function of the dimuon
		     transverse momentum.}
\label{fig:phi_over_rhoomega}
\end{figure}
%
%

The ratio 
$N_\phi/(N_\rho+N_\omega)$=$BR(\phi \rightarrow \mu \mu) \sigma_\phi/
[BR(\rho \rightarrow \mu \mu) \sigma_\rho + BR(\omega \rightarrow \mu \mu) \sigma_\omega]$, 
corrected for acceptance and efficiency, was calculated for $1 < \pt < 5$~GeV/$c$,
giving $0.416 \pm 0.032 \mathrm{(stat.)} \pm 0.004 \mathrm{(syst.)}$.
Systematic uncertainties are due to the normalizations of 
$\omega\rightarrow \mu\mu\pi^0$, $\eta'\rightarrow \mu\mu\gamma$ and combinatorial background.
The corresponding ratio is calculated with PYTHIA and PHOJET.
All the predictions underestimate the measured ratio, as reported in Table~\ref{tab:compModels}. 
The $\pt$ dependence of this ratio is shown in Fig.~\ref{fig:phi_over_rhoomega}.
The Perugia-0, Perugia-11 and D6T tunes systematically underestimate this ratio, 
while PHOJET correctly reproduces the data for $\pt>3$~GeV/$c$, and ATLAS-CSC
is in agreement with the measurement for $\pt>1.5$~GeV/$c$. 

In order to extract the $\omega$ cross section, the $\rho$ and $\omega$ contributions 
must be disentangled, leaving the $\rho$ normalization as an additional free pa\-ra\-me\-ter 
in the fit to the dimuon mass spectrum. The result of the fit 
for $1 < \pt < 5$~GeV/$c$ gives 
$\sigma_\rho / \sigma_\omega = 1.15 \pm 0.20 \mathrm{(stat)} \pm 0.12 \mathrm{(syst)}$,
in agreement with model predictions, as shown in Table~\ref{tab:compModels}.
The systematic uncertainty was evaluated changing the normalizations of the 
$\eta' \rightarrow \mu \mu \gamma$ and $\omega \rightarrow \mu \mu \pi^0$ 
according to the uncertainties in their branching ratios and the 
background level by $\pm 10\%$, which corresponds to twice 
the uncertainty in the normalization. 
The $\omega$ production cross section, calculated from this ratio, is 
$\sigma_\omega (1<\pt<5~{\rm GeV}/c,2.5<y<4) = 5.28 \pm 0.54 \mathrm{(stat)} \pm 0.50 \mathrm{(syst)~mb}$. 
This value is in agreement with the Perugia-0 PYTHIA tune, while the other tunes and 
PHOJET overestimate the $\omega$ cross section, as shown in Table~\ref{tab:compModels}. 

\begin{figure}[tb]
	\centering
	\includegraphics[width=0.60 \columnwidth]{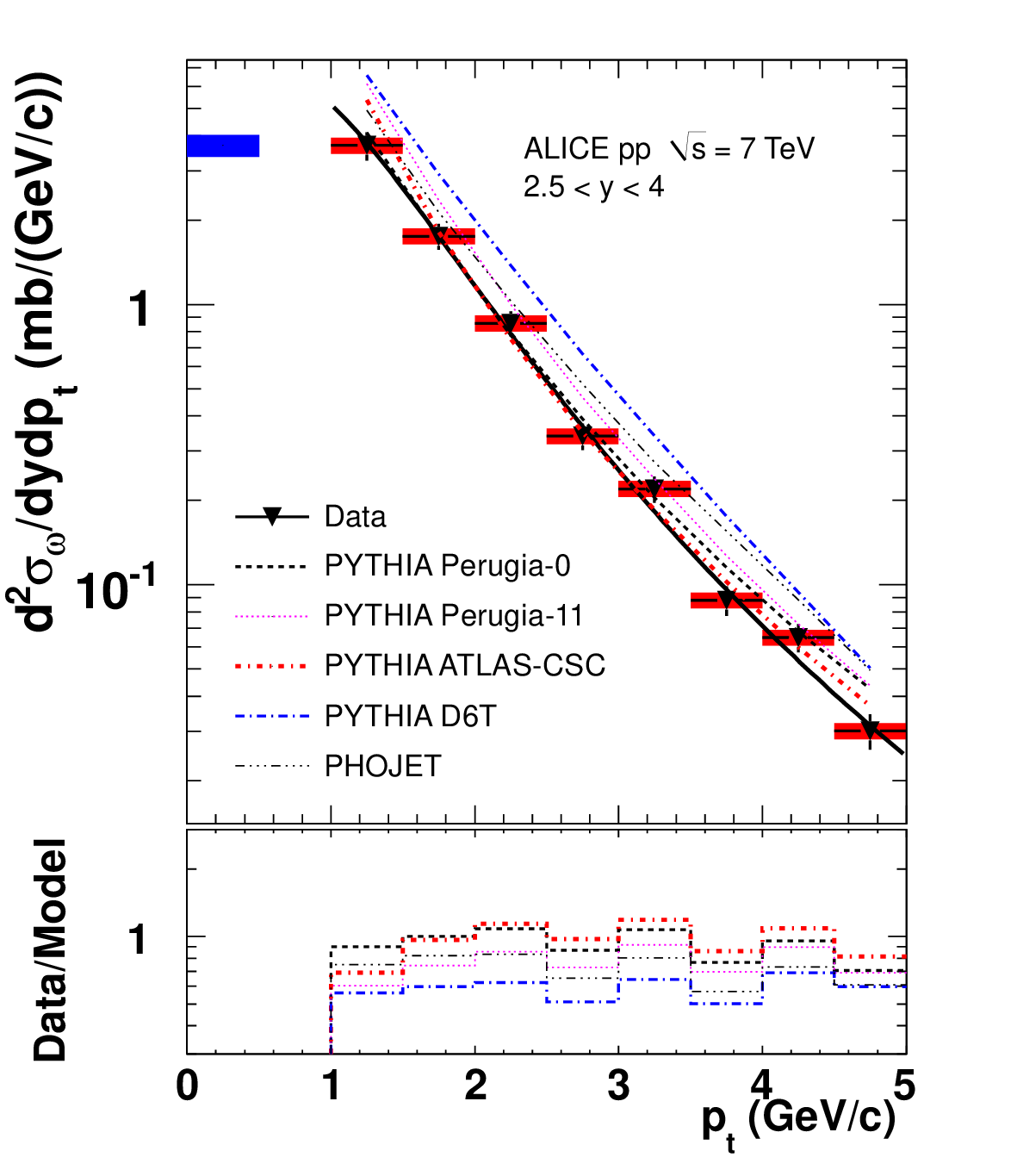}
	\caption{(Color online) Top: Inclusive differential $\omega$ production cross section
	${\rm d}^2\sigma_\omega/{\rm d}y {\rm d}\pt$ for $2.5 < y < 4$. 
	The error bars represent the quadratic sum of the statistical and systematic 
	uncertainties, the red boxes the 
	point-to-point uncorrelated systematic uncertainty, the blue box on 
	the left the error on normalization. Data are fitted with Eq.~(\ref{powerlaw})
	(solid line) and compared with the Perugia-0, Perugia-11, 
	ATLAS-CSC and D6T PYTHIA tunes and PHOJET.
	Bottom: Ratio between data and models. 	
	}
\label{fig:dsigmaomegadydpt}
\end{figure}

In Fig.~\ref{fig:dsigmaomegadydpt} (top) the $\omega$ differential cross section is shown. 
Numerical values are reported in Table~\ref{tab:xsect}.
A fit of Eq.~(\ref{powerlaw}) (solid line) to the data gives
$p_0=1.44 \pm 0.09$~GeV/$c$ and $n=3.2\pm 0.1$.
As shown in the same figure (bottom), all the PYTHIA tunes
reproduce the $\pt$ slope, while PHOJET gives a slightly harder spectrum. 
 
%
%

%
\begin{table*}
\caption{Measured cross sections and ratios compared to the calculation from 
PYTHIA with several tunes and PHOJET in the range $1<\pt <5$~GeV/$c$, $2.5<y<4$}
{\small
\begin{tabular}{lcccc}
\hline
 ~ & $\sigma_\phi$ (mb)& $\sigma_\omega$ (mb) & ${N_{\phi}} \over {N_{\rho} + N_{\omega}}$ & $\sigma_\rho / \sigma_\omega$ \\
\hline
ALICE $\mu\mu$ measurement & $0.940 \pm 0.084 \pm 0.078$ & $5.28 \pm 0.54 \pm 0.50$ & $0.416 \pm 0.032 \pm 0.004$ & $1.15 \pm 0.20 \pm 0.12$ \\
PYTHIA/Perugia-0  & 0.50 & 5.60 & 0.22 & 1.03 \\
PYTHIA/Perugia-11 & 0.62 & 7.81 & 0.20 & 1.03 \\
PYTHIA/ATLAS-CSC  & 0.91 & 6.50 & 0.35 & 1.05 \\
PYTHIA/D6T        & 1.12 & 9.15 & 0.30 & 1.04 \\
PHOJET            & 0.87 & 6.89 & 0.30 & 1.08 \\
\hline
\end{tabular}
}
\label{tab:compModels}
\end{table*}
%
%
\begin{table*}
\caption{$\phi$ and $\omega$ differential cross sections for $2.5 < y < 4$. 
Statistical, bin-to-bin uncorrelated and correlated systematic errors are reported.}
{\small
\begin{tabular}{ccc}
\hline
$\pt$ (GeV/$c$) & ${\rm d}^2\sigma_\phi/{\rm d}y {\rm d}\pt$ (mb/(GeV/$c$)) & ${\rm d}^2\sigma_\omega/{\rm d}y {\rm d}\pt$ (mb/(GeV/$c$)) \\
\hline
$[1,1.5]$ & $0.695   \pm 0.079  \pm 0.046   \pm 0.053  $ & $ 3.69   \pm 0.35   \pm 0.24   \pm 0.32 $ \\
$[1.5,2]$ & $0.268   \pm 0.032  \pm 0.018   \pm 0.020  $ & $ 1.75   \pm 0.15   \pm 0.12   \pm 0.15 $ \\
$[2,2.5]$ & $0.147   \pm 0.014  \pm 0.010   \pm 0.011  $ & $ 0.857  \pm 0.069  \pm 0.057  \pm 0.075 $ \\
$[2.5,3]$ & $0.0665  \pm 0.0074 \pm 0.0044  \pm 0.0051 $ & $ 0.339  \pm 0.029  \pm 0.022  \pm 0.030 $ \\
$[3,3.5]$ & $0.0403  \pm 0.0044 \pm 0.0027  \pm 0.0031 $ & $ 0.220  \pm 0.019  \pm 0.011  \pm 0.019 $ \\
$[3.5,4]$ & $0.0169  \pm 0.0031 \pm 0.0011  \pm 0.0013 $ & $ 0.0880 \pm 0.0088 \pm 0.0058 \pm 0.0077 $ \\
$[4,4.5]$ & $0.0131  \pm 0.0022 \pm 0.0009  \pm 0.0010 $ & $ 0.0648 \pm 0.0062 \pm 0.0043 \pm 0.0056 $ \\
$[4.5,5]$ & $0.0069  \pm 0.0017 \pm 0.0005  \pm 0.0005 $ & $ 0.0301 \pm 0.0039 \pm 0.0020 \pm 0.0026 $ \\
\hline
\end{tabular}
}
\label{tab:xsect}
\end{table*}
%

%
%
\section{Conclusions}
Vector meson production in pp collisions at $\sqrt{s}=7$~TeV was measured 
through the dimuon decay channel in $2.5 < y <4$ and $\pt>1~\mathrm{GeV}/c$. 
The inclusive $\phi$ production cross section 
$\sigma_\phi(1<\pt<5~{\rm GeV}/c,2.5<y<4)=0.940 \pm  0.084 (\mathrm{stat}) \pm 0.078 \mathrm{(syst)~mb}$
was measured with a sample corresponding to an integrated luminosity  
$\mathcal{L}_{\rm int}=55.7~\mathrm{nb}^{-1}$. Calculations based on PHOJET and PYTHIA 
with the ATLAS-CSC and D6T tunes give results that are in agreement 
with the measurement, while the Perugia-0 and Perugia-11 PYTHIA tunes 
underestimate the cross section by about a factor of 2 and 1.5, respectively. 
The ratio 
$N_\phi/(N_\rho+N_\omega)$, calculated for $1<\pt<5$~GeV/$c$, gives 
$0.416 \pm 0.032 \pm 0.004$. This value is reproduced by PHOJET for $\pt>3$~GeV/$c$, 
and by the ATLAS-CSC tune for $\pt>1.5$~GeV/$c$, while the other tunes underestimate
the ratio in the full range $1<\pt<5~{\rm GeV}/c$. 
By measuring the ratio of the $\rho$ and $\omega$ cross sections, 
$\sigma_\rho / \sigma_\omega = 1.15 \pm 0.20 \mathrm{(stat)} \pm 0.12 \mathrm{(syst)}$,
it was possible to extract the inclusive $\omega$ production cross section 
$\sigma_\omega (1<\pt<5~{\rm GeV}/c,2.5<y<4) = 5.28 \pm 0.54 \mathrm{(stat)} \pm 0.50 \mathrm{(syst)~mb}$. 
While all models correctly reproduce the measured $\sigma_\rho / \sigma_\omega$ ratio, the 
$\omega$ cross section is correctly reproduced only by the Perugia-0 calculation, 
and overestimated by the others. 
The differential production cross sections of $\omega$ and $\phi$ were measured. 
The $\pt$ dependence of the $\phi$ cross section agrees
well with other measurements done in the kaon decay channel. 
The ATLAS-CSC and D6T tunes correctly reproduce the $\phi$ $\pt$ 
spectrum, while the other calculations predict harder spectra. 
PHOJET predicts also a slightly harder $\pt$ spectrum for the $\omega$, while 
PYTHIA provides slopes which are closer to the 
one obtained with this measurement. 
\label{Conclusions}
%
%
%
%
%
\section{Acknowledgements}
The ALICE collaboration would like to thank all its engineers and technicians for their invaluable contributions to the construction of the experiment and the CERN accelerator teams for the outstanding performance of the LHC complex.
\\
The ALICE collaboration acknowledges the following funding agencies for their support in building and
running the ALICE detector:
 \\
Department of Science and Technology, South Africa;
 \\
Calouste Gulbenkian Foundation from Lisbon and Swiss Fonds Kidagan, Armenia;
 \\
Conselho Nacional de Desenvolvimento Cient\'{\i}fico e Tecnol\'{o}gico (CNPq), Financiadora de Estudos e Projetos (FINEP),
Funda\c{c}\~{a}o de Amparo \`{a} Pesquisa do Estado de S\~{a}o Paulo (FAPESP);
 \\
National Natural Science Foundation of China (NSFC), the Chinese Ministry of Education (CMOE)
and the Ministry of Science and Technology of China (MSTC);
 \\
Ministry of Education and Youth of the Czech Republic;
 \\
Danish Natural Science Research Council, the Carlsberg Foundation and the Danish National Research Foundation;
 \\
The European Research Council under the European Community's Seventh Framework Programme;
 \\
Helsinki Institute of Physics and the Academy of Finland;
 \\
French CNRS-IN2P3, the `Region Pays de Loire', `Region Alsace', `Region Auvergne' and CEA, France;
 \\
German BMBF and the Helmholtz Association;
\\
General Secretariat for Research and Technology, Ministry of
Development, Greece;
\\
Hungarian OTKA and National Office for Research and Technology (NKTH);
 \\
Department of Atomic Energy and Department of Science and Technology of the Government of India;
 \\
Istituto Nazionale di Fisica Nucleare (INFN) of Italy;
 \\
MEXT Grant-in-Aid for Specially Promoted Research, Ja\-pan;
 \\
Joint Institute for Nuclear Research, Dubna;
 \\
National Research Foundation of Korea (NRF);
 \\
CONACYT, DGAPA, M\'{e}xico, ALFA-EC and the HELEN Program (High-Energy physics Latin-American--European Network);
 \\
Stichting voor Fundamenteel Onderzoek der Materie (FOM) and the Nederlandse Organisatie voor Wetenschappelijk Onderzoek (NWO), Netherlands;
 \\
Research Council of Norway (NFR);
 \\
Polish Ministry of Science and Higher Education;
 \\
National Authority for Scientific Research - NASR (Autoritatea Na\c{t}ional\u{a} pentru Cercetare \c{S}tiin\c{t}ific\u{a} - ANCS);
 \\
Federal Agency of Science of the Ministry of Education and Science of Russian Federation, International Science and
Technology Center, Russian Academy of Sciences, Russian Federal Agency of Atomic Energy, Russian Federal Agency for Science and Innovations and CERN-INTAS;
 \\
Ministry of Education of Slovakia;
 \\
CIEMAT, EELA, Ministerio de Educaci\'{o}n y Ciencia of Spain, Xunta de Galicia (Conseller\'{\i}a de Educaci\'{o}n),
CEA\-DEN, Cubaenerg\'{\i}a, Cuba, and IAEA (International Atomic Energy Agency);
 \\
Swedish Reseach Council (VR) and Knut $\&$ Alice Wallenberg Foundation (KAW);
 \\
Ukraine Ministry of Education and Science;
 \\
United Kingdom Science and Technology Facilities Council (STFC);
 \\
The United States Department of Energy, the United States National
Science Foundation, the State of Texas, and the State of Ohio.

\newpage

\appendix
\section{The ALICE Collaboration}
\label{app:collab}

\begingroup
\small
\begin{flushleft}
B.~Abelev\Irefn{org1234}\And
A.~Abrahantes~Quintana\Irefn{org1197}\And
D.~Adamov\'{a}\Irefn{org1283}\And
A.M.~Adare\Irefn{org1260}\And
M.M.~Aggarwal\Irefn{org1157}\And
G.~Aglieri~Rinella\Irefn{org1192}\And
A.G.~Agocs\Irefn{org1143}\And
A.~Agostinelli\Irefn{org1132}\And
S.~Aguilar~Salazar\Irefn{org1247}\And
Z.~Ahammed\Irefn{org1225}
% ...
\renewcommand\labelenumi{\textsuperscript{\theenumi}~}
\section*{Affiliation notes}
\renewcommand\theenumi{\roman{enumi}}
\begin{Authlist}
\item \Adef{0}Deceased
\item \Adef{Dipartimento di Fisica dell'Universita, Udine, Italy}Also at: Dipartimento di Fisica dell'Universita, Udine, Italy
\item \Adef{M.V.Lomonosov Moscow State University, D.V.Skobeltsyn Institute of Nuclear Physics, Moscow, Russia}Also at: M.V.Lomonosov Moscow State University, D.V.Skobeltsyn Institute of Nuclear Physics, Moscow, Russia
\item \Adef{Institute of Nuclear Sciences, Belgrade, Serbia}Also at: "Vin\v{c}a" Institute of Nuclear Sciences, Belgrade, Serbia
\end{Authlist}
\section*{Collaboration Institutes}
\renewcommand\theenumi{\arabic{enumi}~}
\begin{Authlist}
\item \Idef{org1279}Benem\'{e}rita Universidad Aut\'{o}noma de Puebla, Puebla, Mexico
\item \Idef{org1220}Bogolyubov Institute for Theoretical Physics, Kiev, Ukraine
\item \Idef{org1262}Budker Institute for Nuclear Physics, Novosibirsk, Russia
\item \Idef{org1292}California Polytechnic State University, San Luis Obispo, California, United States
\item \Idef{org14939}Centre de Calcul de l'IN2P3, Villeurbanne, France
\item \Idef{org1197}Centro de Aplicaciones Tecnol\'{o}gicas y Desarrollo Nuclear (CEADEN), Havana, Cuba
\item \Idef{org1242}Centro de Investigaciones Energ\'{e}ticas Medioambientales y Tecnol\'{o}gicas (CIEMAT), Madrid, Spain
\item \Idef{org1244}Centro de Investigaci\'{o}n y de Estudios Avanzados (CINVESTAV), Mexico City and M\'{e}rida, Mexico
\item \Idef{org1335}Centro Fermi -- Centro Studi e Ricerche e Museo Storico della Fisica ``Enrico Fermi'', Rome, Italy
\item \Idef{org17347}Chicago State University, Chicago, United States
\item \Idef{org1118}China Institute of Atomic Energy, Beijing, China
\item \Idef{org1288}Commissariat \`{a} l'Energie Atomique, IRFU, Saclay, France
\item \Idef{org1294}Departamento de F\'{\i}sica de Part\'{\i}culas and IGFAE, Universidad de Santiago de Compostela, Santiago de Compostela, Spain
\item \Idef{org1106}Department of Physics Aligarh Muslim University, Aligarh, India
\item \Idef{org1121}Department of Physics and Technology, University of Bergen, Bergen, Norway
\item \Idef{org1162}Department of Physics, Ohio State University, Columbus, Ohio, United States
\item \Idef{org1300}Department of Physics, Sejong University, Seoul, South Korea
\item \Idef{org1268}Department of Physics, University of Oslo, Oslo, Norway
\item \Idef{org1132}Dipartimento di Fisica dell'Universit\`{a} and Sezione INFN, Bologna, Italy
\item \Idef{org1315}Dipartimento di Fisica dell'Universit\`{a} and Sezione INFN, Trieste, Italy
\item \Idef{org1145}Dipartimento di Fisica dell'Universit\`{a} and Sezione INFN, Cagliari, Italy
\item \Idef{org1270}Dipartimento di Fisica dell'Universit\`{a} and Sezione INFN, Padova, Italy
\item \Idef{org1285}Dipartimento di Fisica dell'Universit\`{a} `La Sapienza' and Sezione INFN, Rome, Italy
\item \Idef{org1154}Dipartimento di Fisica e Astronomia dell'Universit\`{a} and Sezione INFN, Catania, Italy
\item \Idef{org1290}Dipartimento di Fisica `E.R.~Caianiello' dell'Universit\`{a} and Gruppo Collegato INFN, Salerno, Italy
\item \Idef{org1312}Dipartimento di Fisica Sperimentale dell'Universit\`{a} and Sezione INFN, Turin, Italy
\item \Idef{org1103}Dipartimento di Scienze e Tecnologie Avanzate dell'Universit\`{a} del Piemonte Orientale and Gruppo Collegato INFN, Alessandria, Italy
\item \Idef{org1114}Dipartimento Interateneo di Fisica `M.~Merlin' and Sezione INFN, Bari, Italy
\item \Idef{org1237}Division of Experimental High Energy Physics, University of Lund, Lund, Sweden
\item \Idef{org1192}European Organization for Nuclear Research (CERN), Geneva, Switzerland
\item \Idef{org1227}Fachhochschule K\"{o}ln, K\"{o}ln, Germany
\item \Idef{org1122}Faculty of Engineering, Bergen University College, Bergen, Norway
\item \Idef{org1136}Faculty of Mathematics, Physics and Informatics, Comenius University, Bratislava, Slovakia
\item \Idef{org1274}Faculty of Nuclear Sciences and Physical Engineering, Czech Technical University in Prague, Prague, Czech Republic
\item \Idef{org1229}Faculty of Science, P.J.~\v{S}af\'{a}rik University, Ko\v{s}ice, Slovakia
\item \Idef{org1184}Frankfurt Institute for Advanced Studies, Johann Wolfgang Goethe-Universit\"{a}t Frankfurt, Frankfurt, Germany
\item \Idef{org1215}Gangneung-Wonju National University, Gangneung, South Korea
\item \Idef{org1212}Helsinki Institute of Physics (HIP) and University of Jyv\"{a}skyl\"{a}, Jyv\"{a}skyl\"{a}, Finland
\item \Idef{org1203}Hiroshima University, Hiroshima, Japan
\item \Idef{org1329}Hua-Zhong Normal University, Wuhan, China
\item \Idef{org1254}Indian Institute of Technology, Mumbai, India
\item \Idef{org1266}Institut de Physique Nucl\'{e}aire d'Orsay (IPNO), Universit\'{e} Paris-Sud, CNRS-IN2P3, Orsay, France
\item \Idef{org1277}Institute for High Energy Physics, Protvino, Russia
\item \Idef{org1249}Institute for Nuclear Research, Academy of Sciences, Moscow, Russia
\item \Idef{org1320}Nikhef, National Institute for Subatomic Physics and Institute for Subatomic Physics of Utrecht University, Utrecht, Netherlands
\item \Idef{org1250}Institute for Theoretical and Experimental Physics, Moscow, Russia
\item \Idef{org1230}Institute of Experimental Physics, Slovak Academy of Sciences, Ko\v{s}ice, Slovakia
\item \Idef{org1127}Institute of Physics, Bhubaneswar, India
\item \Idef{org1275}Institute of Physics, Academy of Sciences of the Czech Republic, Prague, Czech Republic
\item \Idef{org1139}Institute of Space Sciences (ISS), Bucharest, Romania
\item \Idef{org27399}Institut f\"{u}r Informatik, Johann Wolfgang Goethe-Universit\"{a}t Frankfurt, Frankfurt, Germany
\item \Idef{org1185}Institut f\"{u}r Kernphysik, Johann Wolfgang Goethe-Universit\"{a}t Frankfurt, Frankfurt, Germany
\item \Idef{org1177}Institut f\"{u}r Kernphysik, Technische Universit\"{a}t Darmstadt, Darmstadt, Germany
\item \Idef{org1256}Institut f\"{u}r Kernphysik, Westf\"{a}lische Wilhelms-Universit\"{a}t M\"{u}nster, M\"{u}nster, Germany
\item \Idef{org1246}Instituto de Ciencias Nucleares, Universidad Nacional Aut\'{o}noma de M\'{e}xico, Mexico City, Mexico
\item \Idef{org1247}Instituto de F\'{\i}sica, Universidad Nacional Aut\'{o}noma de M\'{e}xico, Mexico City, Mexico
\item \Idef{org23333}Institut of Theoretical Physics, University of Wroclaw
\item \Idef{org1308}Institut Pluridisciplinaire Hubert Curien (IPHC), Universit\'{e} de Strasbourg, CNRS-IN2P3, Strasbourg, France
\item \Idef{org1182}Joint Institute for Nuclear Research (JINR), Dubna, Russia
\item \Idef{org1143}KFKI Research Institute for Particle and Nuclear Physics, Hungarian Academy of Sciences, Budapest, Hungary
\item \Idef{org18995}Kharkiv Institute of Physics and Technology (KIPT), National Academy of Sciences of Ukraine (NASU), Kharkov, Ukraine
\item \Idef{org1199}Kirchhoff-Institut f\"{u}r Physik, Ruprecht-Karls-Universit\"{a}t Heidelberg, Heidelberg, Germany
\item \Idef{org20954}Korea Institute of Science and Technology Information
\item \Idef{org1160}Laboratoire de Physique Corpusculaire (LPC), Clermont Universit\'{e}, Universit\'{e} Blaise Pascal, CNRS--IN2P3, Clermont-Ferrand, France
\item \Idef{org1194}Laboratoire de Physique Subatomique et de Cosmologie (LPSC), Universit\'{e} Joseph Fourier, CNRS-IN2P3, Institut Polytechnique de Grenoble, Grenoble, France
\item \Idef{org1187}Laboratori Nazionali di Frascati, INFN, Frascati, Italy
\item \Idef{org1232}Laboratori Nazionali di Legnaro, INFN, Legnaro, Italy
\item \Idef{org1125}Lawrence Berkeley National Laboratory, Berkeley, California, United States
\item \Idef{org1234}Lawrence Livermore National Laboratory, Livermore, California, United States
\item \Idef{org1251}Moscow Engineering Physics Institute, Moscow, Russia
\item \Idef{org1140}National Institute for Physics and Nuclear Engineering, Bucharest, Romania
\item \Idef{org1165}Niels Bohr Institute, University of Copenhagen, Copenhagen, Denmark
\item \Idef{org1109}Nikhef, National Institute for Subatomic Physics, Amsterdam, Netherlands
\item \Idef{org1283}Nuclear Physics Institute, Academy of Sciences of the Czech Republic, \v{R}e\v{z} u Prahy, Czech Republic
\item \Idef{org1264}Oak Ridge National Laboratory, Oak Ridge, Tennessee, United States
\item \Idef{org1189}Petersburg Nuclear Physics Institute, Gatchina, Russia
\item \Idef{org1170}Physics Department, Creighton University, Omaha, Nebraska, United States
\item \Idef{org1157}Physics Department, Panjab University, Chandigarh, India
\item \Idef{org1112}Physics Department, University of Athens, Athens, Greece
\item \Idef{org1152}Physics Department, University of Cape Town, iThemba LABS, Cape Town, South Africa
\item \Idef{org1209}Physics Department, University of Jammu, Jammu, India
\item \Idef{org1207}Physics Department, University of Rajasthan, Jaipur, India
\item \Idef{org1200}Physikalisches Institut, Ruprecht-Karls-Universit\"{a}t Heidelberg, Heidelberg, Germany
\item \Idef{org1325}Purdue University, West Lafayette, Indiana, United States
\item \Idef{org1281}Pusan National University, Pusan, South Korea
\item \Idef{org1176}Research Division and ExtreMe Matter Institute EMMI, GSI Helmholtzzentrum f\"ur Schwerionenforschung, Darmstadt, Germany
\item \Idef{org1334}Rudjer Bo\v{s}kovi\'{c} Institute, Zagreb, Croatia
\item \Idef{org1298}Russian Federal Nuclear Center (VNIIEF), Sarov, Russia
\item \Idef{org1252}Russian Research Centre Kurchatov Institute, Moscow, Russia
\item \Idef{org1224}Saha Institute of Nuclear Physics, Kolkata, India
\item \Idef{org1130}School of Physics and Astronomy, University of Birmingham, Birmingham, United Kingdom
\item \Idef{org1338}Secci\'{o}n F\'{\i}sica, Departamento de Ciencias, Pontificia Universidad Cat\'{o}lica del Per\'{u}, Lima, Peru
\item \Idef{org1146}Sezione INFN, Cagliari, Italy
\item \Idef{org1115}Sezione INFN, Bari, Italy
\item \Idef{org1313}Sezione INFN, Turin, Italy
\item \Idef{org1133}Sezione INFN, Bologna, Italy
\item \Idef{org1155}Sezione INFN, Catania, Italy
\item \Idef{org1316}Sezione INFN, Trieste, Italy
\item \Idef{org1286}Sezione INFN, Rome, Italy
\item \Idef{org1271}Sezione INFN, Padova, Italy
\item \Idef{org1322}Soltan Institute for Nuclear Studies, Warsaw, Poland
\item \Idef{org1258}SUBATECH, Ecole des Mines de Nantes, Universit\'{e} de Nantes, CNRS-IN2P3, Nantes, France
\item \Idef{org1304}Technical University of Split FESB, Split, Croatia
\item \Idef{org11602}test~institute
\item \Idef{org1168}The Henryk Niewodniczanski Institute of Nuclear Physics, Polish Academy of Sciences, Cracow, Poland
\item \Idef{org17361}The University of Texas at Austin, Physics Department, Austin, TX, United States
\item \Idef{org1173}Universidad Aut\'{o}noma de Sinaloa, Culiac\'{a}n, Mexico
\item \Idef{org1296}Universidade de S\~{a}o Paulo (USP), S\~{a}o Paulo, Brazil
\item \Idef{org1149}Universidade Estadual de Campinas (UNICAMP), Campinas, Brazil
\item \Idef{org1239}Universit\'{e} de Lyon, Universit\'{e} Lyon 1, CNRS/IN2P3, IPN-Lyon, Villeurbanne, France
\item \Idef{org1205}University of Houston, Houston, Texas, United States
\item \Idef{org20371}University of Technology and Austrian Academy of Sciences, Vienna, Austria
\item \Idef{org1222}University of Tennessee, Knoxville, Tennessee, United States
\item \Idef{org1310}University of Tokyo, Tokyo, Japan
\item \Idef{org1318}University of Tsukuba, Tsukuba, Japan
\item \Idef{org21360}Eberhard Karls Universit\"{a}t T\"{u}bingen, T\"{u}bingen, Germany
\item \Idef{org1225}Variable Energy Cyclotron Centre, Kolkata, India
\item \Idef{org1306}V.~Fock Institute for Physics, St. Petersburg State University, St. Petersburg, Russia
\item \Idef{org1323}Warsaw University of Technology, Warsaw, Poland
\item \Idef{org1179}Wayne State University, Detroit, Michigan, United States
\item \Idef{org1260}Yale University, New Haven, Connecticut, United States
\item \Idef{org1332}Yerevan Physics Institute, Yerevan, Armenia
\item \Idef{org15649}Yildiz Technical University, Istanbul, Turkey
\item \Idef{org1301}Yonsei University, Seoul, South Korea
\item \Idef{org1327}Zentrum f\"{u}r Technologietransfer und Telekommunikation (ZTT), Fachhochschule Worms, Worms, Germany
\end{Authlist}
\endgroup

\end{document}